\documentclass[cup5b]{caps}
\usepackage{graphicx}
\usepackage{amssymb}
\usepackage{ociwsymp2}
\HeadText{C. S. Kochanek and P. L. Schechter}

\newdimen\sa  \newdimen\sb
\def\farcs{\sa=.07em \sb=.03em
     \ifmmode $\rlap{.}$^{\scriptscriptstyle\prime\kern -\sb\prime}$\kern -\sa$
     \else \rlap{.}$^{\scriptscriptstyle\prime\kern -\sb\prime}$\kern -\sa\fi}
\def\farcm{\sa=.08em \sb=.03em
     \ifmmode $\rlap{.}\kern\sa$^{\scriptscriptstyle\prime}$\kern-\sb$
     \else \rlap{.}\kern\sa$^{\scriptscriptstyle\prime}$\kern-\sb\fi}
\def\gtorder{\mathrel{\raise.3ex\hbox{$>$}\mkern-14mu
             \lower0.6ex\hbox{$\sim$}}}
\def\ltorder{\mathrel{\raise.3ex\hbox{$<$}\mkern-14mu
             \lower0.6ex\hbox{$\sim$}}}

\def\vx{\vec{x}}
\def\vu{\vec{\beta}}
\def\vn{\vec{\nabla}}
\def\kbar{\langle\kappa\rangle}
\def\rbar{\langle R \rangle }
\def\dr{\Delta R}
\def\vsis{48\pm3}
\def\vml{71\pm3}

\def\kmsmpc {~km~s$^{-1}$~Mpc$^{-1}$}

\begin{document}

\pagenumbering{arabic}

\author[]{CHRISTOPHER S. KOCHANEK\\Harvard-Smithsonian Center for Astrophysics
\and
PAUL L. SCHECHTER\\Massachusetts Institute of Technology }

\chapter{The Hubble Constant from \\ Gravitational Lens Time Delays}

\begin{abstract}
There are now 10 firm time delay measurements in gravitational lenses.  The
physics of time delays is well understood, and the only important variable
for interpreting the time delays to determine $H_0$ is the mean surface mass 
density $\kbar$ (in units of the critical density for gravitational
lensing) of the lens galaxy at the radius of the lensed images.  More centrally
concentrated mass distributions with lower $\kbar$ predict higher Hubble
constants, with $H_0 \propto 1-\kbar$ to lowest order.  While we cannot
determine $\kbar$ directly given the available data on the current time
delay lenses, we find $H_0=\vsis$~\kmsmpc\ for the isothermal (flat
rotation curve) models, which are our best present estimate for the mass
distributions of the lens galaxies.  Only if we eliminate the dark matter
halo of the lenses and use a constant mass-to-light ratio ($M/L$) model to 
find $H_0 = \vml $~\kmsmpc\ 
is the result consistent with local estimates.  Measurements of time delays in 
better-constrained systems or observations to obtain new constraints on the 
current systems provide a clear path to eliminating the $\kbar$ degeneracy
and making estimates of $H_0$ with smaller uncertainties than are possible
locally.  Independent of the value of $H_0$, the time delay lenses provide
a new and unique probe of the dark matter distributions of galaxies and
clusters because they measure the total (light $+$ dark) matter surface 
density.
\end{abstract}

\section{Introduction}

Fifteen years prior to their discovery in 1979, Refsdal (1964) outlined how
gravitationally lensed quasars might be used to determine the Hubble
constant.  Astronomers have spent the quarter century since their
discovery working out the difficult details not considered in
Refsdal's seminal papers.

The difficulties encountered fall into two broad categories ---
measurement and modeling.  Time delays can be hard to measure if the
fluxes of the images do not vary, or if the images are faint, or if they lie
very close to each other.  Modeling gravitational potentials with a
small number of constraints is likewise difficult, either because the
lens geometry is complex or because the data poorly constrain the 
most important aspects of the gravitational potential.
We will argue that these difficulties are surmountable, both 
in principle and in practice, and
that an effort considerably smaller than that of the {\it HST}\ Hubble
Constant Key Project will yield a considerably smaller uncertainty in
the Hubble constant, $H_0$. 

While the number of systems with measured time delays is small,
their interpretation implies a value for $H_0$, which, given our current 
understanding of the dark matter distributions
of galaxies, is formally inconsistent with that obtained using Cepheids.
The Key Project value of $H_0=72\pm8$~\kmsmpc\
(Freedman et al.~2001) is consistent with the lens data
only if the lens galaxies have significantly less dark matter
than is expected theoretically or has been measured for other early-type
galaxies.  While it is premature to argue for replacing the
local estimates, we hope to persuade the astronomical community
that the time delay result deserves both careful attention 
and further study.

Interpreting time delays requires a model for the gravitational
potential of the lens, and in most cases the uncertainties in 
the model dominate the uncertainty in $H_0$.  Thus, the
main focus of this review will be to explain the dependence of
time delays on gravitational potentials.
We start in \S\ref{sec:basics} by introducing the
time delay method and illustrating the physics of time delays
with a series of simple models.  In \S\ref{sec:theory} we 
review a general mathematical theory of time delays to show
that, for most lenses, the only important parameter of the model
is the mean surface density of the lens at the radius of the images.  In 
\S\ref{sec:clusters} we discuss the effects of the environment
of the lens on time delays.  We review the data on the time
delay lenses in \S\ref{sec:data} and their implications for
the Hubble constant and dark matter in early-type galaxies 
in \S\ref{sec:results}.  The present time delay lenses have
a degeneracy between $H_0$ and the amount of dark matter, so
in \S\ref{sec:profile} we outline several approaches that can
eliminate the degeneracy.  Finally, in \S\ref{sec:future} we
discuss the future of time delays.  Unless otherwise stated, 
we assume a flat, $\Omega_m=0.3$, $\Omega_{\Lambda}=0.7$ cosmological
model.  

\begin{figure*}[t]
\includegraphics[width=1.00\columnwidth,angle=0]{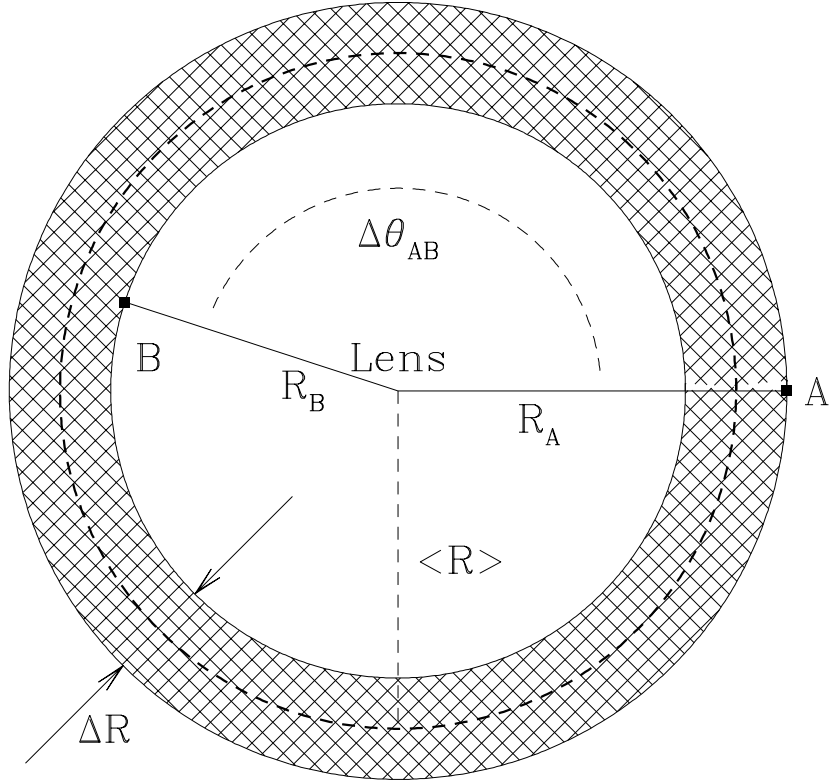}
\vskip 0pt \caption{
Schematic diagram of a two-image time delay lens. The lens lies at the
origin, with two images A and B at radii $R_A$ and $R_B$ from the lens center.
The images define an annulus of average radius 
$\protect{\rbar}=(R_A+R_B)/2$ and width $\protect{\dr}=R_A-R_B$,
and the images subtend an angle $\Delta\theta_{AB}$ relative to the lens 
center.  For a circular lens $\Delta\theta_{AB}=180^\circ$ by symmetry. 
\label{fig:geometry}}
\end{figure*}

\section{Time Delay Basics \label{sec:basics}}

The observations of gravitationally lensed quasars are best understood
in light of Fermat's principle (e.g., Blandford \& Narayan 1986).
Intervening mass between a source and an observer introduces an
effective index of refraction, thereby increasing the light-travel
time.  The competition between this Shapiro delay from the 
gravitational field and the geometric delay due to bending the
ray paths leads to the formation of multiple images at  
the stationary points (minima, maxima, and saddle points) of the 
travel time (for more complete reviews, see Narayan \& Bartelmann 1999 
or Schneider, Ehlers, \& Falco 1992).

As with glass optics, there is a thin-lens approximation that applies 
when the optics are small compared to the distances to the source 
and the observer.  In this approximation, we need only the effective 
potential, $\psi(\vx)=(2/c^2)(D_{ls}/D_s)\int dz \phi $, found
by integrating the 3D potential $\phi$ along the line of sight.
The light-travel time is
\begin{equation}
  \tau\left(\vx\right) = \left[{1+z_l \over c}\right]
  \left[{D_{l} D_{s}\over D_{ls} }\right]
  \left[ {1\over 2} \left(\vx - \vec{\beta}\right)^2 -
  \psi\left(\vx\right)\right],
\end{equation}
where $\vx=(x,y)=R(\cos\theta,\sin \theta)$ and $\vu$ are the angular
positions of the image and the source, $\psi(\vx)$ is the effective potential, 
$(\vx-\vu)^2/2$ is the geometric delay in the small-angle approximation,
$z_l$ is the lens redshift,
and $D_l$, $D_s$, and $D_{ls}$ are angular-diameter distances 
to the lens, to the source, and from the lens to the source,
respectively.  The only dimensioned quantity in the travel time
is a factor of $H_0^{-1}\simeq 10 h^{-1}$~Gyr arising from the
$H_0^{-1}$ scaling of the angular-diameter distances.
 
We observe the images at the extrema of the time delay function, which we find
by setting the gradients with respect to the image positions
equal to zero, $\vn_x \tau = 0$, and finding all the stationary 
points  ($\vx_A$, $\vx_B$, $\cdots$) associated
with a given source position $\vu$.  The local magnification of an
image is determined by the magnification tensor $M_{ij}$, whose
inverse is determined by the second derivatives of the time
delay function, 
\begin{equation}
    M_{ij}^{-1} = \vn_x \vn_x \tau(\vx)
          =
   \left(
   \begin{array}{cc}
      1-\kappa-\gamma\cos 2\theta_\gamma &\gamma\sin 2\theta_\gamma \\
      \gamma\sin 2\theta_\gamma          &1-\kappa+\gamma\cos 2\theta_\gamma 
   \end{array}
   \right),
\end{equation}  
where the convergence $\kappa=\Sigma/\Sigma_c$ is the local surface density 
in units of the critical surface density 
$\Sigma_c = c^2 D_s / 4 \pi G D_l D_{ls}$, 
and $\gamma$ and $\theta_\gamma$ define the local shear field and its 
orientation.    The determinant of the magnification tensor is the net
magnification of the image, but it is a signed quantity depending on 
whether the image has positive (maxima, minima) or negative (saddle points)
parity.  

A simple but surprisingly realistic starting point for modeling lens 
potentials is the singular isothermal sphere (the SIS model) in which
the lens potential is simply
\begin{equation}
\psi(\vx) = b R,
  \qquad\hbox{where}\qquad
  b = 4 \pi {D_{ls} \over D_s} {\sigma^2 \over c^2} 
  = \hbox{1\farcs45} 
    \left( {\sigma \over 225\, \hbox{km s}^{-1} }\right)^2 { D_{ls} \over D_s }
\end{equation}
is a deflection scale determined by geometry and $\sigma$ is the 1D velocity 
dispersion of the lens galaxy.  For $|\vu| < b$, the SIS lens
produces two colinear images at radii $R_A=|\vu|+b$ and $R_B=b-|\vu|$ on
opposite sides of the lens galaxy (as in 
Fig.~\ref{fig:geometry} but with $\Delta\theta_{AB}=180^\circ$).\footnote{
  The deflections produced by the SIS lens are constant, 
  $|\protect{\vx}-\protect{\vu}|=b$,
  so the total image separation is always $2b$.  The outer image is
  brighter than the inner image, with signed magnifications 
  $M_A^{-1}=1-b/R_A > 0 $ (a positive parity minimum) 
  and $M_B^{-1}=1-b/R_B < 0$ (a negative parity saddle point). 
  The model parameters, $b=(R_A+R_B)/2=\protect{\rbar}$ and 
  $|\protect{\vu}|=(R_A-R_B)/2=\protect{\dr}/2$, can be
  determined uniquely from the image positions.  }  
The A image is a minimum of the time delay and leads the 
saddle point, B, with a time delay difference of 
\begin{equation}
  \Delta t_{SIS} = \tau_B - \tau_A =  {1\over 2}
   \left[{1+z_l\over c}\right] \left[{D_{l} D_{s}\over D_{ls} }\right]
     (R_A^2 - R_B^2) .
   \label{eqn:sis}
\end{equation}
Typical time delay differences of months or years are the consequence of
multiplying the $\sim 10h^{-1}$~Gyr total 
propagation times by the square of a very small
angle ($b \approx  3 \times 10^{-6}$ radians so, $R_A^2 \approx 10^{-11}$). 
The SIS model suggests that lens time delay measurements reduce the 
determination of the Hubble constant to a problem of differential
astrometry.  This is almost correct, but we have made two idealizations
in using the SIS model.

The first idealization was to ignore deviations of the radial
(monopole) density profile from that of an SIS with density $\rho \propto r^{-2}$, 
surface density $\Sigma \propto R^{-1}$, and a flat rotation curve.
The SIS is a  special case of a power-law monopole with lens potential
\begin{equation}
\psi(\vx)  = {b^2 \over (3-\eta)} \left({R\over b}\right)^{3-\eta},
\end{equation}
corresponding to a (3D) density distribution with density $\rho \propto r^{-\eta}$,
surface density $\Sigma \propto R^{1-\eta}$, and rotation curve
$\upsilon_c \propto r^{(2-\eta)/2}$.  For $\eta=2$ we recover the SIS model,
and the normalization is chosen so that the scale $b$ is always the Einstein
ring radius.  Models with smaller (larger) $\eta$ have less (more)
centrally concentrated mass distributions and have rising (falling) rotation 
curves.  The limit $\eta \rightarrow 3$ approaches the potential of a point
mass.  By adjusting the scale $b$ and the source position $|\vu|$, we can 
fit the observed positions of two images at radii
$R_A$ and $R_B$ on opposite sides ($\Delta\theta_{AB}=180^\circ$)
of the lens for any value of $\eta$.\footnote{ 
  In theory we have one additional constraint
  because the image flux ratio measures the magnification ratio,
  $f_A/f_B=|M_A|/|M_B|$, and the magnification ratio depends on $\eta$.
  Unfortunately, the systematic errors created by milli- and 
  microlensing make it difficult to use flux ratios as 
  model constraints (see \S\protect{\ref{sec:data}}).}
The expression for the time delay difference can be well approximated by
(Witt, Mao, \& Keeton 2000; Kochanek~2002)
\begin{equation}
  \Delta t(\eta) = \tau_B-\tau_A \simeq
    (\eta - 1) \Delta t_{SIS}
    \left[ 1 - { (2-\eta)^2 \over 12 }
         \left( { \dr \over \rbar } \right)^2 \cdots
    \right],
    \label{eqn:powerlaw}
\end{equation}
where $\rbar=(R_A+R_B)/2\simeq b$ and $\dr=R_A-R_B$ (see Fig.~\ref{fig:geometry}).
While the expansion assumes $\dr/\rbar$ (or $|\vu|$) is small, 
we can usually ignore the higher-order terms.
There are two important lessons from this model.  
\begin{enumerate}
\item Image astrometry of simple two-image and four-image lenses generally
   cannot constrain the radial mass distribution of the lens.
\item More centrally concentrated mass distributions (larger $\eta$) 
   predict longer time delays, resulting in a larger Hubble constant
   for a given time delay measurement.
\end{enumerate}
These problems, which we will address from a different perspective in 
\S\ref{sec:theory}, are the cause of the uncertainties in estimates of $H_0$
from time delays.

The second idealization was to ignore deviations from circular symmetry due to 
either the ellipticity of the lens galaxy or the local tidal gravity field from 
nearby objects.  A very nice analytic example of a lens with angular structure
is a singular isothermal model with {\it arbitrary}
angular structure, where the effective potential is $\psi = b R F(\theta )$,
and $F(\theta)$ is an arbitrary function. The model family includes
the most common lens model, the singular isothermal ellipsoid (SIE).
The time delays for this model family are simply $\Delta t_{SIS}$, 
{\it independent of the angular structure of the lens} 
(Witt et al.~2000)!  This result, while attractive, does not hold in
general, and we will require the results of \S\ref{sec:theory} to
understand the effects of angular structure in the potential. 

\section{Understanding Time Delays: A General Theory \label{sec:theory}}

The need to model the gravitational potential of the lens is the aspect of
interpreting time delays that creates the greatest suspicion.  The most
extreme view is that it renders the project ``hopeless'' because we will 
never be able to guarantee that the models encompass the degrees of freedom 
needed to capture all the systematic uncertainties.  In order to address
these fears we must show that we understand the specific properties of
the gravitational potential determining time delays and then ensure that
our parameterized models include these degrees of freedom.  

The examples we considered in \S\ref{sec:basics} illustrate the basic physics
of time delays, but an extensive catalog of (non)parametric models demonstrating
the same properties may not be convincing to the skeptic.  We will instead
show, using standard mathematical expansions of the potential, 
which properties of the lens galaxy are required to understand time delays 
with accuracies of a few percent.   While we can
understand the results of all models for the time delays of gravitational
lenses based on this simple theory, full numerical models should probably
be used for most detailed, quantitative analyses.   Fortunately,
there are publically available programs for both the parametric and
nonparametric approaches.\footnote{ The {\it gravlens} and {\it lensmodel}
  (Keeton 2003, cfa-www.harvard.edu/$\sim$castles) packages include a very
  broad range of parametric models for the mass distributions of lenses,
  and the {\it PixelLens} package (Williams \& Saha~2000,
  ankh-morpork.maths.qmw.ac.uk/$\sim$saha/astron/lens/pix/) implements
  a nonparametric approach. }
Our analysis uses the geometry
of the schematic lens shown in Figure~\ref{fig:geometry}. The two images define
an annulus bounded by their radii, $R_A$ and $R_B$, and with an interior region 
for $R< R_B$ and an exterior region for $R >R_A$.

The key to understanding time delays comes from Gorenstein, Falco, \& Shapiro
(1988; see also Saha~2000), who showed that the time delay of 
a circular lens depends only on the image positions and the 
{\it surface density $\kappa(R)$ in the annulus between the images}. 
The mass of the interior region is implicit in the image positions
and accurately determined by the astrometry.  From Gauss' law, we
know that the radial distribution of the mass in the interior 
region and the amount or distribution of mass in the exterior 
region is irrelevant.  A useful approximation is
to assume that the surface density in the annulus can be
{\it locally} approximated by a power law $\kappa \propto R^{1-\eta}$
and that the mean surface density in the annulus is 
$\kbar=\langle\Sigma\rangle/\Sigma_c$.  The time delay between
the images is (Kochanek~2002)
\begin{equation}
    \Delta t =  2 \Delta t_{SIS} \left[ 1-\kbar   
        - { 1 - \eta \kbar \over 12 } \left( { \dr \over \rbar } \right)^2
        + O\left( \left( { \dr \over \rbar } \right)^4 \right) \right].
     \label{eqn:monopole}
\end{equation}
Thus, the time delay is largely determined by the average density $\kbar$, with
only modest corrections from the local shape of the surface density 
distribution    even when $\dr/\rbar \simeq 1$.  For example, the second-order 
expansion is exact for an SIS lens ($\kbar=1/2$, $\eta=2$) and reproduces the 
time delay of a point mass lens ($\kbar=0$) to better than 1\% even when
$\dr/\rbar =1$.  This local model also explains the time delay scalings
of the global power-law models we discussed in \S\ref{sec:basics}.  A
$\rho \propto r^{-\eta}$ global power law has surface density 
$\kbar=(3-\eta)/2$ near the Einstein ring, so the
leading term of the time delay is 
$\Delta t = 2 \Delta t_{SIS} (1-\kbar) = (\eta-1)\Delta t_{SIS}$,
just as in Equation~(\ref{eqn:powerlaw}).  

\begin{itemize}
\item 
The time delay is not determined 
by the global structure of the radial density profile but rather by the 
surface density near the Einstein ring.
\end{itemize}

Gorenstein et al.~(1988) considered only circular lenses,
but a multipole expansion allows us to understand the role of angular
structure (Kochanek~2002).   An estimate to the same order as in 
Equation~(\ref{eqn:monopole}) requires only the quadrupole moments of the 
regions interior and exterior to the annulus, provided the strengths 
of the higher-order multipoles of the potential have the same order
of magnitude as for an ellipsoidal density distribution.\footnote{
  If the quadrupole potential, $\psi_2 \propto \cos 2\theta$,
  has dimensionless amplitude 
  $\epsilon_2$, then it produces ray deflections of order $O(\epsilon_2 b)$
  at the Einstein ring of the lens.  In a four-image lens the quadrupole
  deflections are comparable to the thickness of the annulus, so
  $\epsilon_2 \simeq \dr/\rbar$.  In a two-image lens they are smaller than
  the thickness of the annulus, so $\epsilon_2 \ltorder \dr/\rbar$.
  For an ellipsoidal density distribution, the
  $\cos (2m\theta)$ multipole amplitude scales as 
  $\epsilon_{2m} \sim \epsilon_2^m \ltorder (\dr/\rbar)^m$. This allows
  us to ignore the quadrupole density distribution in the annulus and
  all higher-order multipoles.  It is important to remember that
  potentials are much rounder than surface densities [with relative
  amplitudes for a $\cos (m\theta)$ multipole of roughly
  $m^{-2}$:$m^{-1}$:1 for potentials:deflections:densities], so the
  multipoles relevant to time delays converge rapidly even for very
  flat surface density distributions.
  } 
This approximation can fail for the lenses in clusters (see \S\ref{sec:clusters}).
The complete expansion for $\Delta t$ when the two quadrupole
moments have independent amplitudes and orientations is not very
informative.  However, the leading term of the expansion when the two
quadrupole moments are aligned illustrates the role of angular structure.
Given an exterior quadrupole (i.e., an external shear) of amplitude 
$\gamma_{ext}$ and an interior quadrupole of amplitude $\gamma_{int}$
sharing a common axis $\theta_\gamma$, the quadrupole potential is
\begin{equation}
   \psi_2 = { 1 \over 2 } \left( \gamma_{ext} R^2 +
      \gamma_{int} { \rbar^4 \over R^2 } \right)
      \cos 2(\theta-\theta_\gamma)
\end{equation}
if we define the amplitudes at radius $\rbar$.  For images at positions
$R_A(\cos \theta_A, \sin \theta_A)$ and 
$R_B(\cos \theta_B, \sin \theta_B)$ relative to the lens galaxy
(see Fig.~\ref{fig:geometry}),
the leading term of the time delay is
\def\dt12{\Delta\theta_{AB}}
\begin{equation}
    \Delta t \simeq 2 \Delta t_{SIS} (1-\kbar) 
      { \sin^2 (\dt12/2) \over 1 - 4 f_{int} \cos^2 (\dt12/2) },
\end{equation}
where $\dt12 = \theta_A-\theta_B$ and 
$f_{int}=\gamma_{int}/(\gamma_{ext}+\gamma_{int})$ is the 
fraction of the quadrupole due to the interior quadrupole moment
$\gamma_{int}$.  We need not worry about the possibility of a 
singular denominator --- successful global models of the lens
do not allow such configurations.

A two-image lens has too few astrometric constraints to fully
constrain a model with independent, misaligned internal and external
quadrupoles.  Fortunately, when the lensed images lie on opposite
sides of the lens galaxy ($\dt12 \simeq \pi +\delta$, $|\delta| \ll 1$),
the time delay becomes insensitive to the quadrupole structure.
Provided the angular deflections are smaller than the radial 
deflections ($|\delta|\rbar \ltorder \dr$), the leading term of
the time delay reduces to the result for a circular lens, 
$\Delta t \simeq 2 \Delta t_{SIS} (1-\kbar)$.  There is, however,
one limiting case to remember.  If the images and the lens 
are colinear, as in a spherical lens, the component of the shear
aligned with the separation vector acts like a contribution to
the convergence.  In most lenses this would be a modest additional
uncertainty --- in the typical lens these shears must be small, 
the sign of the effect should be nearly random, and it is only
a true degeneracy in the limit that everything is colinear.  

A four-image lens has more astrometric constraints and can 
constrain a model with independent, misaligned internal and external
quadrupoles.  The quadrupole moments of the observed lenses are dominated 
by external shear, with $f_{int} \ltorder 1/4$ unless there is more than
one lens galaxy inside the Einstein ring.  The ability of the
astrometry to constrain $f_{int}$ is important because the delays
depend strongly on $f_{int}$ when the images do not lie on opposite
sides of the galaxy.  If external shears dominate, $f_{int}\simeq 0$ 
and the leading term of the delay becomes 
$\Delta t \simeq 2\Delta t_{SIS}(1-\kbar)\sin^2 \dt12/2$.  
If the model is isothermal,  like the $\psi = r F(\theta)$ models we 
considered in \S\ref{sec:basics}, then $f_{int}=1/4$ and we again
find that the delay is independent of the angle, with 
$\Delta t \simeq 2\Delta t_{SIS}(1-\kbar)$. 
The time delay ratios in a four-image lens are largely determined
by the angular structure and provide a check of the potential model.

In summary, if we want to understand time delays to an accuracy competitive
with studies of the local distance scale (5\%--10\%), the only important
variable is the surface density $\kbar$ of the lens in the annulus between
the images.  When models based on the same data for the time delay and
the image positions predict different values for $H_0$, the differences
can always be understood as the consequence of different choices for
$\kbar$.  In parametric models $\kbar$ is adjusted by changing the
central concentration of the lens (i.e., like $\eta$ in the global
power-law models), and in the nonparametric models of Williams
\& Saha~(2000) it is adjusted directly.  The expansion models of
Zhao \& Qin (2003a,b) mix aspects of both approaches.  

\section{Lenses Within Clusters \label{sec:clusters}}

Most galaxies are not isolated, and many early-type lens galaxies are
members of groups or clusters, so we need to consider the effects of
the local environment on the time delays.  Weak perturbations are
easily understood since they will simply be additional contributions
to the surface density ($\kbar$) and the external shear/quadrupole 
($\gamma_{ext}$) we discussed in \S\ref{sec:theory}.  In this 
section we focus on the consequences of large perturbations.  

As a first approximation we can assume that a nearby cluster (or galaxy)
can be modeled by an SIS potential,  $\Psi_c(\vx) = B|\vec{x} - \vec{x}_c|$,
where $B$ is the Einstein radius of the cluster and 
$\vec{x}_c=R_C(\cos \theta_c,\sin \theta_c)$ is the 
position of the cluster relative to the primary lens.  We can
understand its effects by expanding the potential as a 
series in $R/R_c$, dropping constant and linear terms that
have no observable consequences, to find that
\begin{equation}
\Psi_c \simeq  { 1 \over 4 } {B \over R_c} R^2 
  - { 1 \over 4 } {B \over R_c} R^2 \cos 2(\theta-\theta_c)
     + O\left( { B \over R_c^2 } R^3 \right). 
   \label{eqn:cluster}
\end{equation}
The first term has
the form $(1/2) \kappa_c R^2$, which is the potential of a uniform sheet whose
surface density $\kappa_c = B/2 R_c$ is that of the cluster at the lens
center.  The second term has the form $(1/2) \gamma_c R^2 \cos 2(\theta-\theta_c)$, 
which is the (external) tidal shear $\gamma_c= B/2 R_c$ that would be produced by 
putting all the cluster mass inside a ring of radius $R_c$ at the 
cluster center.  
All realistic lens models
need to incorporate a tidal shear term due to objects near the lens or 
along the line of sight (Keeton, Kochanek, \& Seljak 1997), but as we discussed in 
\S\ref{sec:theory} the shear does not lead to significant ambiguities 
in the time delay estimates.  Usually the local shear cannot be associated
with a particular object unless it is quite strong ($\gamma_c \approx 0.1$).\footnote{
There is a small random component of $\kappa$ contributed by material
along the line of sight (Barkana 1996). This introduces small uncertainties 
in the $H_0$ estimates for individual lenses (an rms convergence of 
$0.01-0.05$, 
depending on the source redshift), but is an unimportant source of uncertainty 
in estimates from ensembles of lenses because it is a random variable that 
averages to zero.}  

The problems with nearby objects arise when the convergence $\kappa_c$ 
becomes large because of a global degeneracy known as the {\it mass-sheet degeneracy} 
(Falco, Gorenstein, \& Shapiro 1985).  If we have a model predicting a time delay $\Delta t_0$ 
and then add a sheet of constant surface density $\kappa_c$, then the time delay 
is changed to $(1-\kappa_c)\Delta t_0$ without changing the image positions,
flux ratios, or time delay ratios.  Its effects can be understood from
\S\ref{sec:theory} as a contribution to the annular surface density with
$\kbar=\kappa_c$ and $\eta=1$.  The parameters of the lens, in particular
the mass scale $b$, are also rescaled by factors of $1-\kappa_c$, so the  
degeneracy can be broken if there is an independent
mass estimate for either the cluster or the galaxy.\footnote{
  For the cluster this can be done using weak lensing (e.g., Fischer et al. 1997
  in Q0957+561), cluster galaxy velocity dispersions (e.g., Angonin-Willaime,
 Soucail, \& Vanderriest 1994
  for Q0957+561, Hjorth et al. 2002
  for RXJ0911+0551) or X-ray temperatures/luminosities (e.g., Morgan et al. 2001
  for RXJ0911+0551 or Chartas et al.~2002 for Q0957+561).  For the lens
  galaxy this can be done with stellar dynamics (Romanowsky \& Kochanek 1999
  for Q0957+561 and PG1115+080, Treu \& Koopmans 2002b for PG1115+080).
  The accuracy of these methods is uncertain at present because each
  suffers from its own systematic uncertainties.  When the lens is in
  the outskirts of a cluster, as in RXJ0911+0551, it is probably 
  reasonable to assume that $\kappa_c \leq \gamma_c$, as most mass 
  distributions are more centrally concentrated than isothermal.}
When the convergence is due to an object like a cluster, there is a strong 
correlation between the convergence $\kappa_c$ and the shear $\gamma_c$ that is
controlled by the density distribution of the cluster (for our 
isothermal model $\kappa_c=\gamma_c$). 
In most circumstances, neglecting the extra surface density coming
from nearby objects (galaxies, groups, clusters) leads to
an overestimate of the Hubble constant because these objects
all have $\kappa_c >0$.

If the cluster is sufficiently close, then we cannot ignore the higher-order 
perturbations in the expansion of Equation~(\ref{eqn:cluster}). 
They are quantitatively important when they produce deflections at 
the Einstein ring radius $b$ of the primary lens, $B(b/R_c)^2$, that are larger
than the astrometric uncertainties.  Because these uncertainties are
small, the higher-order terms quickly become important.  If they are
important but ignored in the models, the results can be very misleading.

\begin{table}
  \begin{center}
  \caption{Time Delay Measurements}
  \begin{tabular}{lccccc}
\hline               
System       &$N_{im}$&$\Delta t$ (days)  &Astrometry  &Model      &Ref. \\
\hline
HE1104--1805 &2       &$161\pm 7$         &$+$    &``simple''         & 1 \\
PG1115+080   &4       &$ 25\pm 2$         &$+$    &``simple''         & 2 \\
SBS1520+530  &2       &$130\pm 3$         &$+$    &``simple''         & 3 \\
B1600+434    &2       &$ 51\pm 2$         &$+/-$  &``simple''         & 4 \\
HE2149--2745 &2       &$103\pm12$         &$+$    &``simple''         & 5 \\
\hline
RXJ0911+0551 &4       &$146\pm 4$         &$+$    & cluster/satellite & 6 \\
Q0957+561    &2       &$417\pm 3$         &$+$    & cluster           & 7 \\ 
B1608+656    &4       &$ 77\pm 2$         &$+/-$  & satellite         & 8 \\
\hline
B0218+357    &2       &$10.5\pm0.2$       &$-$    &``simple''         & 9 \\
PKS1830--211 &2       &$26 \pm 4$         &$-$    &``simple''         & 10 \\
\hline               
B1422+231    &4       &($  8 \pm  3)$     &$+$    &``simple''         & 11 \\
\hline               
  \end{tabular}
  \end{center}
  \label{tab:delays}
  \noindent
   $N_{im}$ is the number of images. 
   $\Delta t$ is the longest of the measured delays and its 1$\sigma$ 
     error; delays in parenthesis require further confirmation.  
   The ``Astrometry'' column indicates the quality of the astrometric data for 
   the system: $+$ (good), $+/-$ (some problems), $-$ (serious problems). 
   The ``Model'' column indicates the type of model needed to interpret the
      delays.  ``Simple'' lenses can be modeled as a single primary lens
      galaxy in a perturbing tidal field.  More complex models are needed 
      if there is a satellite galaxy inside the Einstein ring (``satellite'') 
      of the primary lens galaxy, or if the primary lens belongs to a
      cluster.
    References: 
    (1) Ofek \& Maoz 2003, also see Gil-Merino, Wistozki, \& Wambsganss 2002,
        Pelt, Refsdal, \& Stabell 2002, and Schechter et al. 2002;
    (2) Barkana 1997, based on Schechter et al. 1997; 
    (3) Burud et al. 2002b;       
    (4) Burud et al. 2000, also Koopmans et al. 2000; 
    (5) Burud et al. 2002a; 
    (6) Hjorth et al. 2002; 
    (7) Kundi\'c et al. 1997, also Schild \& Thomson 1997 and 
        Haarsma et al. 1999; 
    (8) Fassnacht et al. 2002; 
    (9) Biggs et al. 1999, also Cohen et al. 2000;
    (10) Lovell et al. 1998;
    (11) Patnaik \& Narasimha 2001.
\end{table}

\begin{figure*}[t]
\centering \leavevmode
\includegraphics[width=0.45\columnwidth,angle=0]{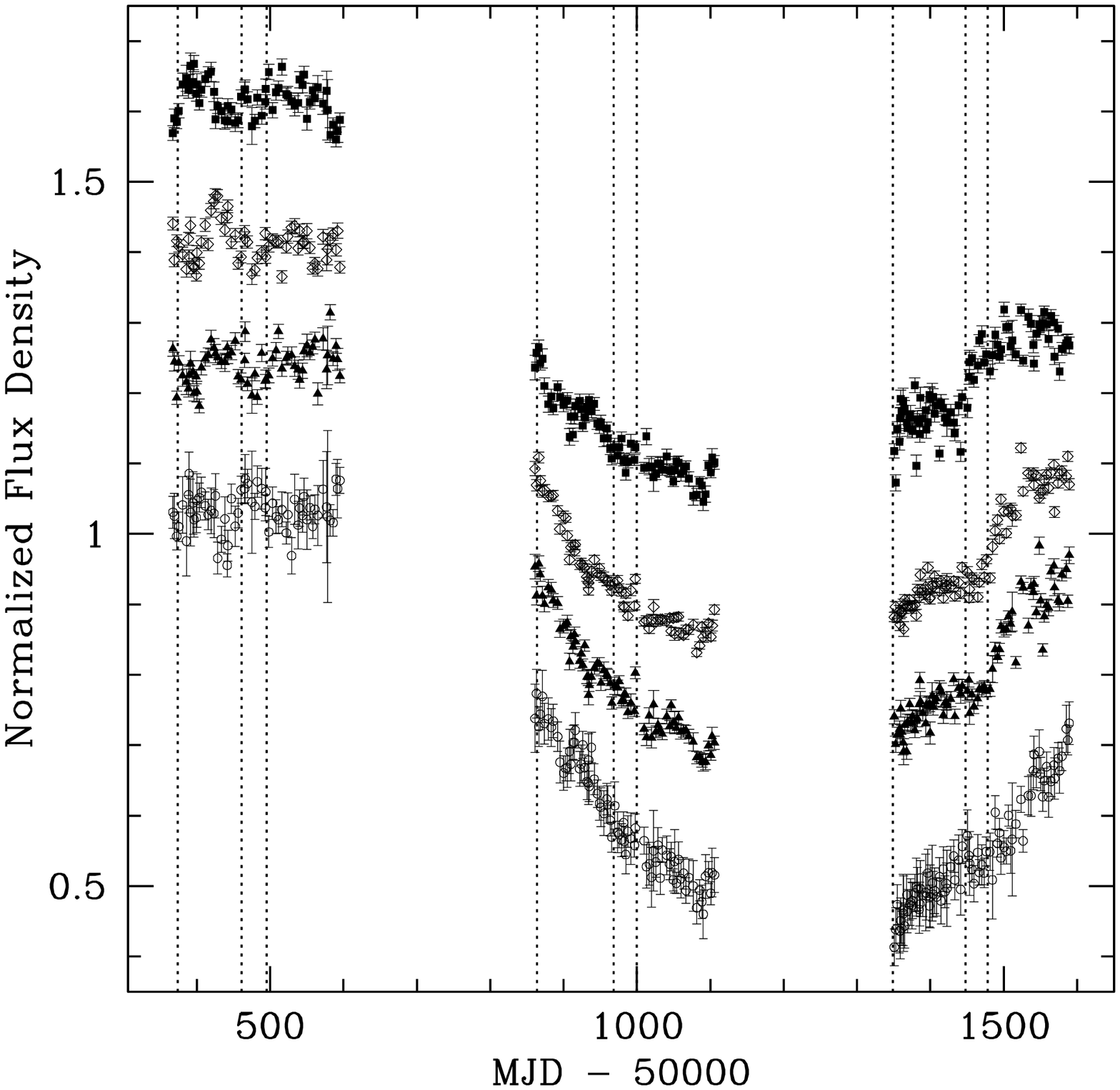} \hfil
\includegraphics[width=0.45\columnwidth,angle=0]{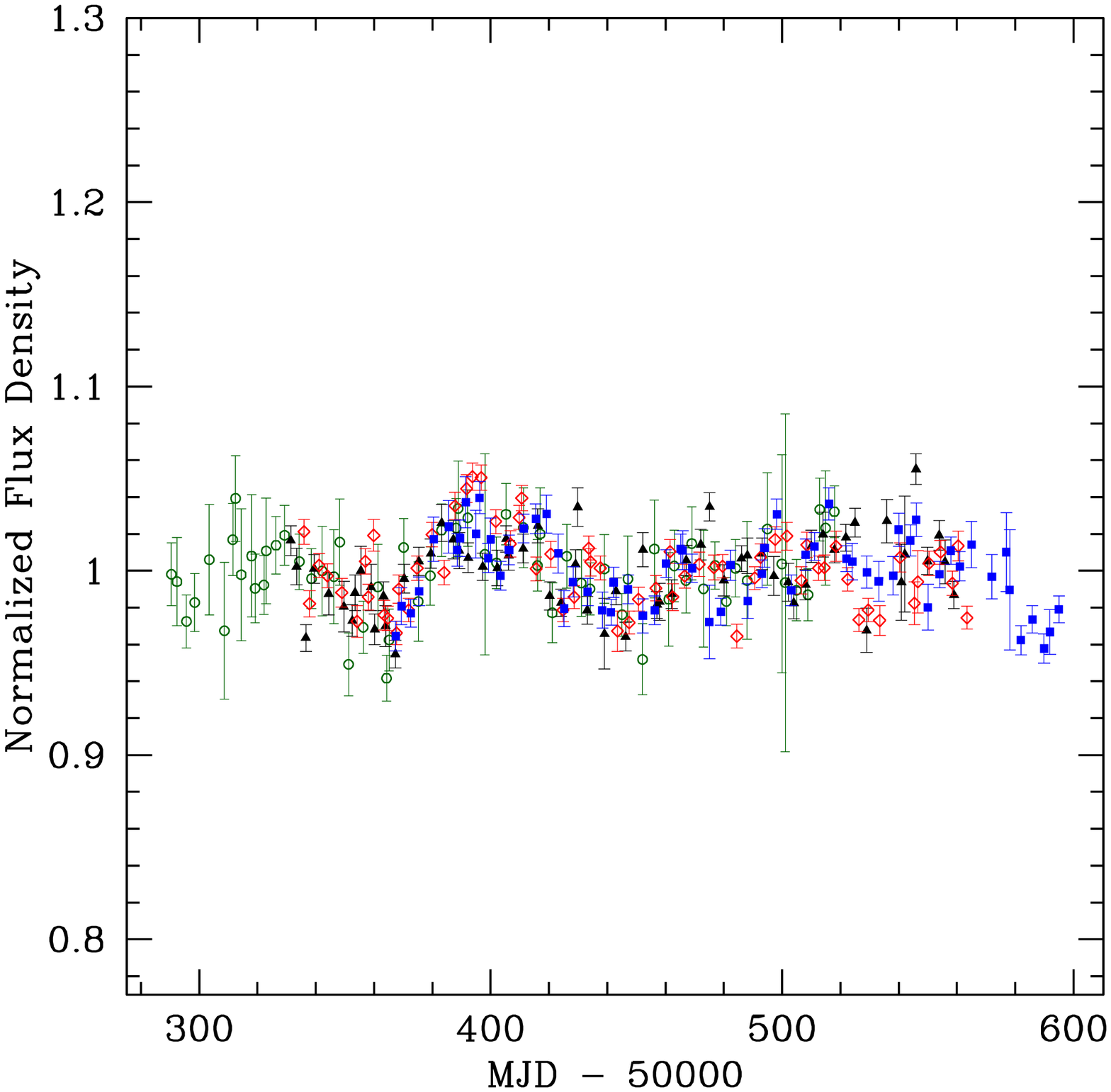}
\vskip 0pt \caption{
VLA monitoring data for the four-image lens B1608+656.  The left panel shows 
(from top to bottom) the normalized light curves for the B (filled squares), A 
(open diamonds), C (filled triangles) and D (open circles) images as a 
function of the mean Julian day.  The right panel shows the composite light 
curve for the first monitoring season after cross correlating the light curves 
to determine the time delays ($\Delta t_{AB}=31.5\pm1.5$, 
$\Delta t_{CB}=36.0\pm1.5$ and $\Delta t_{DB}=77.0\pm1.5$~days) and the flux 
ratios.  (From Fassnacht et al.~2002.)
\label{fig:lcurve}}
\end{figure*}

\section{Observing Time Delays and Time Delay Lenses \label{sec:data}}

The first time delay measurement, for the gravitational lens Q0957+561,
was reported in 1984 (Florentin-Nielsen~1984). 
Unfortunately, a controversy then developed between a short 
delay ($\simeq 1.1$~years, Schild \& Cholfin~1986; Vanderriest et
al.~1989) and a long delay ($\simeq 1.5$~years,
Press, Rybicki, \& Hewitt 1992a,b), which was finally
settled in favor of the short delay only after 5 more years of 
effort (Kundi\'c et al.~1997; also Schild \& Thomson~1997
and Haarsma et al.~1999).  Factors contributing to
the intervening difficulties included the small amplitude of the
variations, systematic effects, which, with hindsight, appear to be due to
microlensing and scheduling difficulties (both technical and
sociological).  

While the long-running controversy over Q0957+561 led to poor publicity for
the measurement of time delays, it allowed the community to 
come to an understanding of the systematic problems in measuring time delays, 
and to develop a broad range of methods for reliably determining time delays 
from typical data.  Only
the sociological problem of conducting large monitoring projects remains
as an impediment to the measurement of time delays in large numbers. 
Even these are slowly being overcome, with the result that the last
five years have seen the publication of time delays in 11 systems (see
Table 1.1). 
 
The basic procedures for measuring a time delay are simple.  A monitoring
campaign must produce light curves for the individual lensed images that 
are well sampled compared to the time delays.  During this period, the
source quasar in the lens must have measurable brightness fluctuations
on time scales shorter than the monitoring period.
The resulting light curves are cross correlated by one or more methods
to measure the delays and their uncertainties  
(e.g., Press et al.~1992a,b; Beskin \& Oknyanskij~1995;
 Pelt et al.~1996; references in Table~ 1.1).
Care must be taken because there can be sources of 
uncorrelated variability between the images due to systematic errors
in the photometry and real effects such as microlensing of the individual
images (e.g., Koopmans et al. 2000; Burud et al. 2002b; Schechter et 
al.~2003).  Figure~\ref{fig:lcurve} shows an example, the
beautiful light curves from the radio lens B1608+656 by Fassnacht
et al.~(2002), where the variations of all four lensed images
have been traced for over three years.  One of the 11 systems,
B1422+231, is limited by systematic uncertainties in the delay
measurements. The brand new time delay for HE1104--1805 (Ofek \& Maoz 2003)
is probably accurate, but has yet to be interpreted in detail. 

We want to have uncertainties in the time delay measurements that 
are unimportant for the estimates of $H_0$.  For the present, 
uncertainties of order 3\%--5\% are adequate (so improved delays
are still needed for PG1115+080, HE2149--2745, and PKS1830--211).
In a four-image lens we can measure three independent time delays,
and the dimensionless ratios of these delays provide
additional constraints on the lens models (see \S\ref{sec:theory}).
These ratios are well measured in B1608+656 (Fassnacht et al.~2002),
poorly measured in PG1115+080 (Barkana~1997; Schechter et al.~1997;
Chartas 2003) and unmeasured in either RXJ0911+0551
or B1422+231.  Using the time delay lenses as very precise 
probes of $H_0$, dark matter and cosmology will eventually require
still smaller delay uncertainties ($\sim 1\%$).  Once a delay is
known to 5\%, it is relatively easy to reduce the uncertainties
further because we can accurately predict when flux variations 
will appear in the other images and need to be monitored.

The expression for the time delay in an SIS lens (Eqn.~\ref{eqn:sis}) 
reveals the other data that are necessary to interpret time delays.  
First, the source and lens redshifts
are needed to compute the distance factors that set the scale of the
time delays.  Fortunately, we know both redshifts for all 11 systems 
in Table~1.1.  The dependence of the angular-diameter distances 
on the cosmological model is unimportant until our total uncertainties 
approach 5\% (see \S\ref{sec:future}). 
Second, we require accurate relative positions for the images and the
lens galaxy.  These uncertainties are always dominated by the position
of the lens galaxy relative to the images.
For most of the lenses in Table~1.1, observations with radio 
interferometers (VLA, Merlin, VLBA) and {\it HST}\ have measured the
relative positions of the images and lenses to accuracies
$\ltorder 0\farcs005$.  Sufficiently deep {\it HST}\ images can obtain
the necessary data for almost any lens, but dust in the lens
galaxy (as seen in B1600+434 and B1608+656) can limit the accuracy 
of the measurement even in a very deep image.  For 
B0218+357 and PKS1830--211, however, the position of the lens galaxy
relative to the images is not known to sufficient precision or is 
disputed (see L\'ehar et al.~2000; Courbin et al.~2002; Winn et al.~2002).

In practice, we fit models of the gravitational potential constrained
by the available data on the image and lens positions, the relative
image fluxes, and the relative time delays.  When imposing these 
constraints, it is important to realize that lens galaxies are not
perfectly smooth.  They contain both low-mass satellites and stars that perturb 
the gravitational potential.  The time delays themselves are completely unaffected by
these substructures.  However, as we take derivatives of the potential
to determine the ray deflections or the magnification, the sensitivity
to substructures in the lens galaxy grows.  Models of substructure in
cold dark matter (CDM) halos predict that the substructure produces
random perturbations of approximately $0\farcs001$ in the image positions
(see Metcalf \& Madau~2001; Dalal \& Kochanek~2002).  We should not
impose tighter astrometric constraints than this limit.
A more serious problem is that substructure, whether
satellites (``millilensing'') or stars (``microlensing''), significantly
affect image fluxes with amplitudes that depend on the image magnification
and parity (see, e.g.,  Wozniak et al.~2000; Burud et al.~2002b;
Dalal \& Kochanek~2002; Schechter et al.~2003 or Schechter \& Wambsganss~2002).
Once the flux errors are enlarged to the 30\% level of these systematic errors, 
they provide little leverage for discriminating between models.

We can also divide the systems by the complexity of the
required lens model.  We define eight of the lenses as ``simple,'' in the
sense that the available data suggests that a model consisting of a
single primary lens in a perturbing shear (tidal gravity) field should
be an adequate representation of the gravitational potential.  In some
of these cases, an external potential representing a nearby galaxy or
parent group will improve the fits, but the differences between the
tidal model and the more complicated perturbing potential are small
(see \S\ref{sec:clusters}).
We include the quotation marks because the classification is based 
on an impression of the systems from the available data and models. 
While we cannot guarantee that a system is simple, we can easily
recognize two complications that will require more complex models.

The first complication is
that some primary lenses have less massive satellite galaxies inside
or near their Einstein rings.  This includes two of the time delay
lenses, RXJ0911+0551 and B1608+656.  RXJ0911+0551 could simply be 
a projection effect, since neither lens galaxy shows irregular 
isophotes.  Here the implication for models may simply be the need
to include all the parameters (mass, position, ellipticity $\cdots$)
required to describe the second lens galaxy, and with more parameters
we would expect greater uncertainties in $H_0$.
In B1608+656, however, the lens galaxies show the 
heavily disturbed isophotes typical of galaxies undergoing a disruptive 
interaction.  How one
should model such a system is unclear. If there was once dark matter 
associated with each of the galaxies, how is it distributed now?  Is it still 
associated with the individual galaxies?  Has it settled into an equilibrium
configuration?  While B1608+656 can be well fit with standard lens
models (Fassnacht et al.~2002), these complications have yet to be explored. 

The second complication occurs when the primary lens is a member of a 
more massive (X-ray) cluster, as in the time delay lenses RXJ0911+0551 
(Morgan et al.~2001) and Q0957+561 (Chartas et al.~2002).  The cluster
model is critical to interpreting these systems (see \S\ref{sec:clusters}).  
The cluster surface density at the position of the lens ($\kappa_c \gtorder 0.2$)
leads to large corrections to the time delay estimates and the higher-order 
perturbations are crucial to obtaining a good model.  For example, models
in which the Q0957+561 cluster was treated simply as an external 
shear are grossly incorrect (see the review of Q0957+561
models in Keeton et al.~2000).  In addition to the uncertainties in
the cluster model itself, we must also decide how to include and
model the other cluster galaxies near the primary lens.  Thus,
lenses in clusters have many reasonable degrees of freedom beyond those
of the ``simple'' lenses.

\section{Results: The Hubble Constant and Dark Matter \label{sec:results}}

With our understanding of the theory and observations of the lenses
we will now explore their implications for $H_0$.  We focus on
the ``simple'' lenses PG1115+080, SBS1520+530, B1600+434, and 
HE2149--2745.  We only comment on the interpretation of the HE1104--1805
delay because the measurement is too recent to have been interpreted
carefully.  We will briefly discuss the more complicated systems
RXJ0911+0551, Q0957+561, and B1608+656, and we will not discuss 
the systems with problematic time delays or astrometry.  

The most common, simple, realistic model of a lens consists of a singular
isothermal ellipsoid (SIE) in an external (tidal) shear field (Keeton et al. 1997).
The model has 7 parameters (the lens position, mass, ellipticity, major axis
orientation for the SIE, and the shear amplitude and orientation).
It has many degrees of freedom associated with the angular
structure of the potential, but the radial structure is fixed with
$\kbar\simeq 1/2$.  For comparison, a two-image (four-image) lens
supplies 5 (13) constraints on any model of the potential: 2 (6) from the
relative positions of the images, 1 (3) from the flux ratios of the images,
0 (2) from the inter-image time delay ratios, and 2 from the lens position.
With the addition of extra components (satellites/clusters) for the
more complex lenses, this basic model provides a good fit to all the
time delay lenses except Q0957+561.  Although a naive counting of the
degrees of freedom ($N_{dof}=-2$ and $6$, respectively) suggests that
estimates of $H_0$ would be underconstrained for two-image lenses and
overconstrained for four-image lenses, the uncertainties are actually
dominated by those of the time delay measurements and the astrometry in
both cases.  This is what we expect from \S\ref{sec:theory} --- the
model has no degrees of freedom that change $\kbar$ or $\eta$, so there
will be little contribution to the uncertainties in $H_0$ from the
model for the potential.

\begin{figure*}[t]
\includegraphics[width=0.80\columnwidth,angle=0]{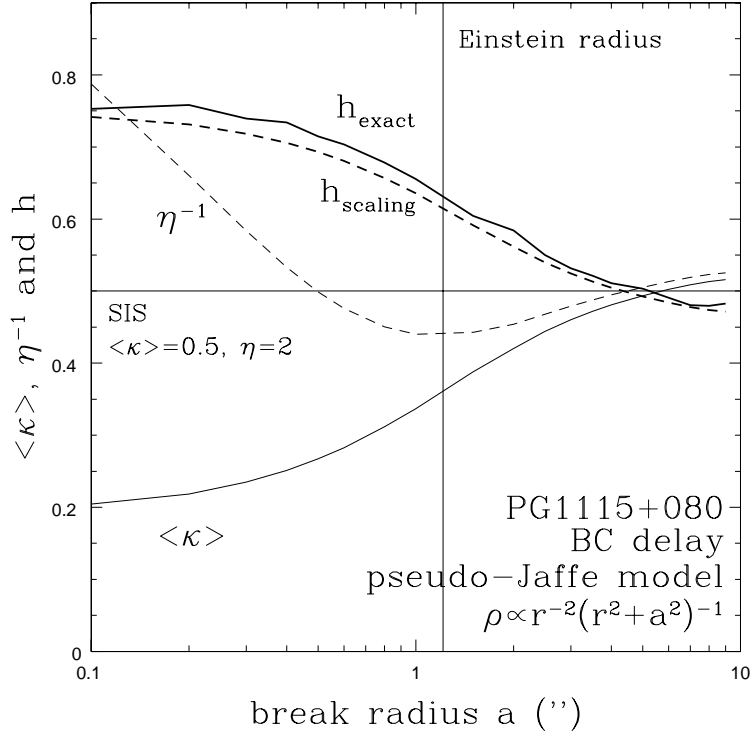}
\vskip 0pt \caption{
$H_0$ estimates for PG1115+080.  The lens galaxy is modeled as a ellipsoidal
pseudo-Jaffe model, $\rho \propto r^{-2}(r^2+a^2)^{-1}$, and the nearby group
is modeled as an SIS.  As the break radius $a \rightarrow \infty$ the
pseudo-Jaffe model becomes an SIS model, and as the break radius
$a\rightarrow 0$ it becomes a point mass.  The heavy solid curve ($h_{exact}$)
shows the dependence of $H_0$ on the break radius for the exact, nonlinear
fits of the model to the PG1115+080 data.  The heavy dashed curve
($h_{scaling}$) is the value found using our simple theory
(\S\protect{\ref{sec:theory}}) of time delays.  The agreement of the exact and
scaling solutions is typical. The light solid line shows the average surface
density $\langle\kappa\rangle$ in the annulus between the images, and the
light dashed line shows the {\it inverse} of the logarithmic slope $\eta$ in
the annulus.  For an SIS model we would have $\langle\kappa\rangle=1/2$ and
$\eta^{-1}=1/2$, as shown by the horizontal line.  When the break radius is
large compared to the Einstein radius (indicated by the vertical line), the
surface density is slightly higher and the slope is slightly shallower than
for the SIS model because of the added surface density from the group.  As we
make the lens galaxy more compact by reducing the break radius, the surface
density decreases and the slope becomes steeper, leading to a rise in $H_0$.
As the galaxy becomes very compact, the surface density near the Einstein ring
is dominated by the group rather than the galaxy, so the surface density
approaches a constant and the logarithmic slope approaches the value
corresponding to a constant density sheet ($\eta=1$).
\label{fig:paul}}
\end{figure*}

If we use a model that includes parameters to control 
the radial density profile (i.e., $\kbar$), 
for example by adding a halo truncation radius $a$ to the SIS profile 
[the pseudo-Jaffe model, $\rho \propto r^{-2} (r^2+a^2)^{-1}$; e.g., Impey
et al.~1998; Burud et al.~2002a],\footnote{This is simply an example.
  The same behavior would be seen for any other parametric model in which
  the radial density profile can be adjusted. } 
then we
find the expected correlation between $a$ and $H_0$ --- as we make the
halo more concentrated (smaller $a$), the estimate of $H_0$ rises
from the value for the SIS profile ($\kbar=1/2$ as $a\rightarrow\infty$)
to the value for a  point mass ($\kbar=0$ as $a\rightarrow 0$),
with the fastest changes occurring when $a$ is similar
to the Einstein radius of the lens. We show an example of such
a model for PG1115+080 in Figure~\ref{fig:paul}.  This case is somewhat
more complicated than a pure pseudo-Jaffe model because there is an 
additional contribution to the surface
density from the group to which the lens galaxy belongs. 
 As long as the structure of the radial density profile is 
fixed (constant $a$), the uncertainties are again dominated by the
uncertainties in the time delay.  Unfortunately, 
the goodness
of fit, $\chi^2(a)$, shows too little dependence on $a$ to determine
$H_0$ uniquely.  In general, two-image lenses have too
few constraints, and the extra constraints supplied by a four-image
lens constrain the angular structure rather than the
radial structure of the potential.
This basic problem holds for all
existing models of the current sample of time delay lenses.

\begin{figure*}[t]
\includegraphics[width=1.00\columnwidth,angle=0]{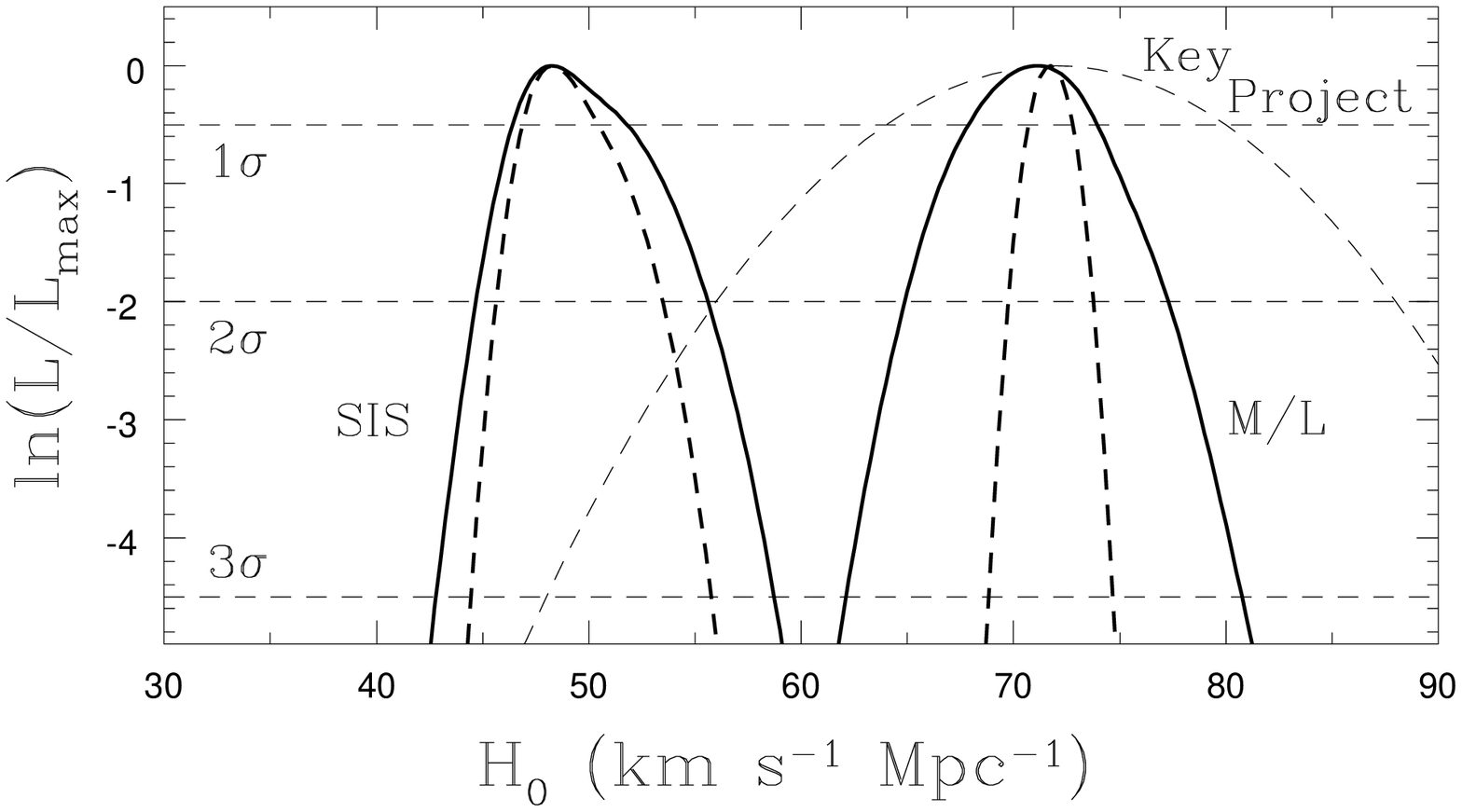}
\vskip 0pt \caption{
$H_0$ likelihood distributions.  The curves show the joint likelihood
functions for $H_0$ using the four simple lenses PG1115+080, SBS1520+530,
B1600+434, and HE2149--2745 and assuming either an SIS model (high
$\protect{\kbar}$, flat rotation curve) or a constant $M/L$ model (low
$\protect{\kbar}$, declining rotation curve).  The heavy dashed curves show
the consequence of including the X-ray time delay for PG1115+080 from Chartas
(2003) in the models. The light dashed curve shows a Gaussian model for the
Key Project result that $H_0=72\pm8$~\kmsmpc.
\label{fig:problem}}
\end{figure*}

The inability of the present time delay lenses to directly constrain the
radial density profile is the major problem for using them to determine
$H_0$. Fortunately, it is a consequence of the available data on the current 
sample rather than a fundamental limitation, as we discuss in the next section 
(\S\ref{sec:profile}).  
It is, however, a simple trade-off -- models with less dark matter 
(lower $\kbar$, more centrally concentrated densities) produce higher 
Hubble constants than those with more dark matter.  
We do have some theoretical limits on the value of $\kbar$. In particular, we
can be confident that the surface density is bounded by 
two limiting models.  The mass distribution should not be more compact
than the luminosity distribution, so a constant mass-to-light ratio ($M/L$)
model should set a lower limit on $\kbar \gtorder \kbar_{M/L} \simeq 0.2$, and
an upper limit on estimates of $H_0$.   We are also confident 
that the typical 
lens should not have a rising rotation curve at 1--2 optical effective radii from 
the center of the lens galaxy.  
Thus, the SIS model is probably the least concentrated reasonable 
model, setting an upper bound on $\kbar \ltorder \kbar_{SIS}=1/2$,  and a 
lower limit on estimates of $H_0$.  Figure~\ref{fig:problem} 
shows joint estimates of $H_0$ from the four simple lenses for these two 
limiting mass distributions (Kochanek~2003b).  The results for the 
individual lenses are mutually consistent and are unchanged by the
new $0.149\pm0.004$~day delay for the A$_1$-A$_2$ images in PG1115+080
(Chartas 2003).  For galaxies with isothermal profiles we
find $H_0=\vsis $~\kmsmpc, and for galaxies with constant $M/L$ 
we find $H_0=\vml $~\kmsmpc.  While our best prior estimate for the
mass distribution is the isothermal profile (see \S\ref{sec:profile}),
the lens galaxies would have to have constant $M/L$ to match 
Key Project estimate of $H_0=72 \pm 8$~\kmsmpc\  (Freedman et al.~2001). 

The difference between these two limits is entirely explained by the
differences in $\kbar$ and $\eta$ between the SIS and constant
$M/L$ models.  In fact, it is possible to reduce the $H_0$ estimates
for each simple lens to an approximation formula,
$H_0 = A(1-\kbar) + B \kbar (\eta -1)$. The coefficients, 
$A$ and $|B| \approx A/10$, are derived from the image positions using 
the simple theory from \S\ref{sec:theory}.
These approximations reproduce numerical results using ellipsoidal
lens models to accuracies of $3$~\kmsmpc\  (Kochanek 2002). 
For example, in Figure~\ref{fig:paul} we also show the estimate of
$H_0$ computed based on the simple theory of \S\ref{sec:theory} and the
annular surface density ($\kbar$) and slope ($\eta$) of the numerical models.  
The agreement
with the full numerical solutions is excellent, even though the
numerical models include both the ellipsoidal lens galaxy and
a group.  No matter what the mass distribution is, the five lenses PG1115+080, 
SBS1520+530, B1600+434, PKS1830--211,\footnote{
PKS1830--211 is included based on the Winn et al.~(2002) model of the
{\it HST}\ imaging data as a single lens galaxy.  Courbin et al.~(2002) prefer 
an interpretation with multiple lens galaxies which would invalidate the 
analysis.}
and HE2149--2745 have very similar dark matter halos.  For a fixed slope 
$\eta$, the five systems are consistent with a common value for the surface 
density of
\begin{equation}
   \kbar = 1 - 1.07 h + 0.14 (\eta-1)(1-h) \pm 0.04
\end{equation}
where $H_0=100h$~\kmsmpc\  and there is an upper limit of 
$\sigma_\kappa \ltorder 0.07$ on the intrinsic scatter of $\kbar$.  Thus,
time delay lenses provide a new window into the structure and homogeneity
of dark matter halos, regardless of the actual value of $H_0$.

\begin{figure*}[t]
\centering \leavevmode 
\includegraphics[width=0.45\columnwidth,angle=0]{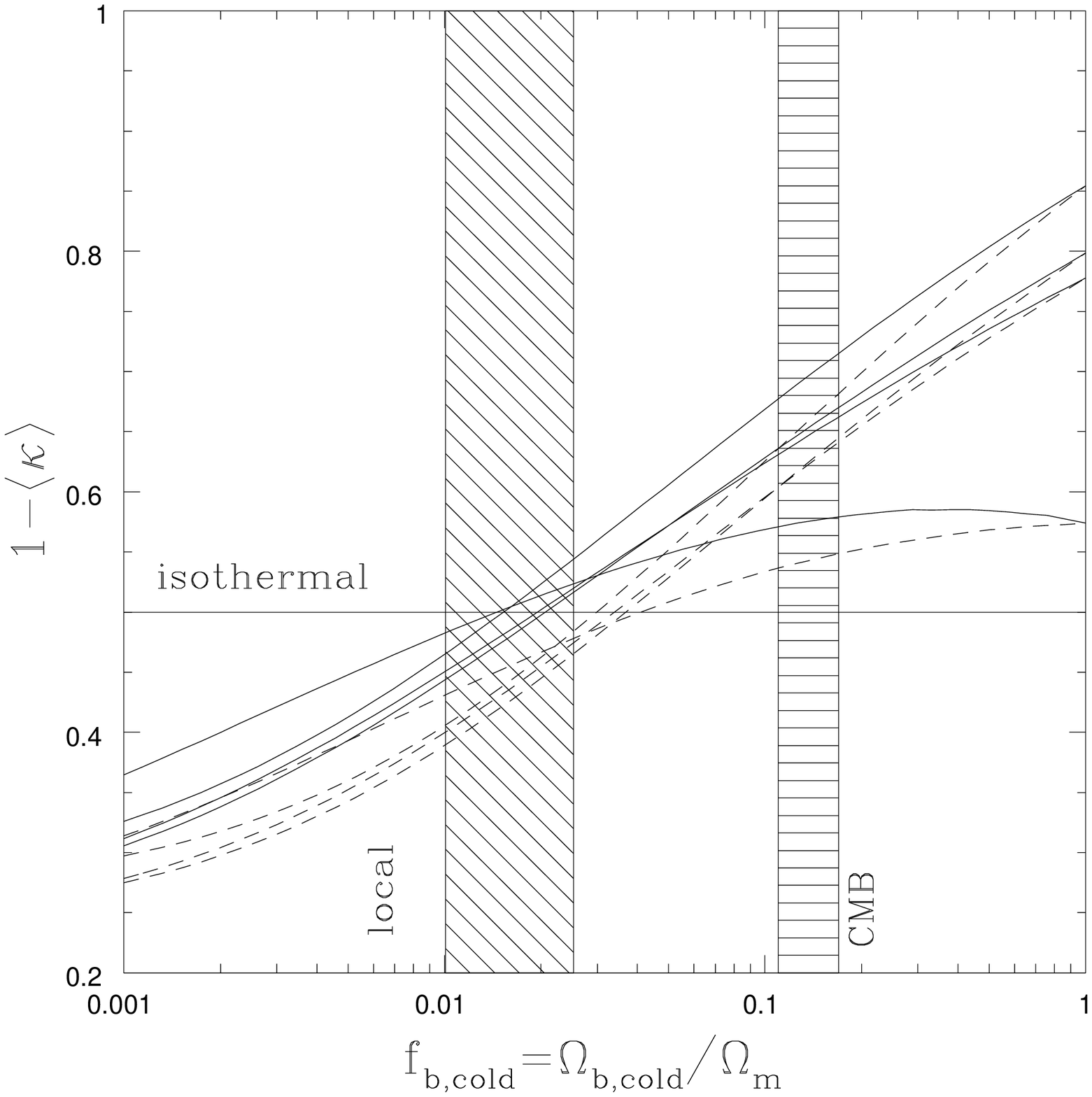} \hfil
\includegraphics[width=0.45\columnwidth,angle=0]{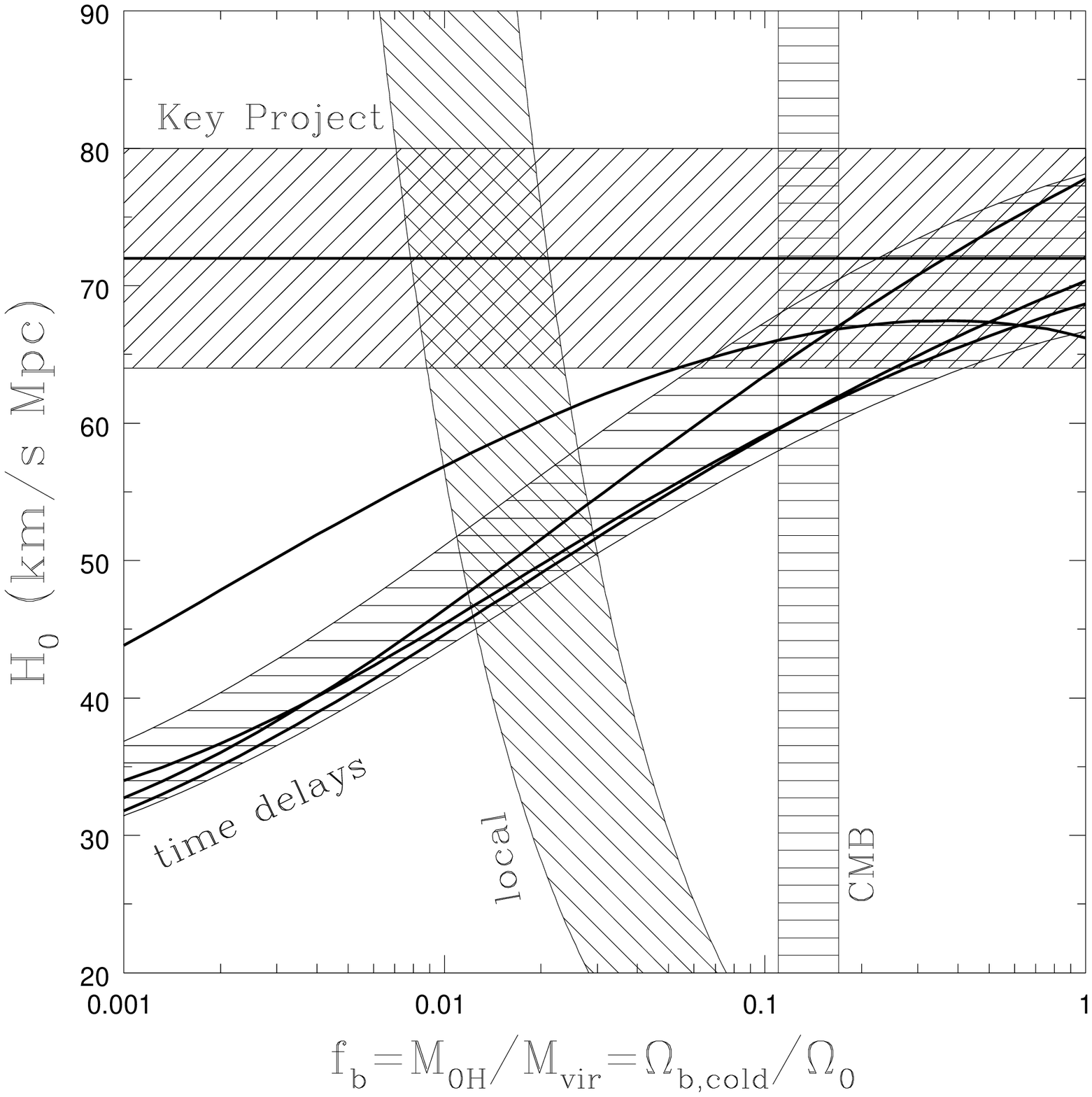}
\vskip 0pt \caption{
$H_0$ in CDM halo models.  The left panel shows $1-\kbar$ for the ``simple''
lenses (PG1115+080, SBS1520+530, B1600+434, and HE2149--2745) as a function of
the cold baryon fraction $f_{b,cold}$.  The solid (dashed) curves include
(exclude) the adiabatic compression of the dark matter by the baryons. The
horizontal line shows the value for an SIS potential.  The right panel shows
the resulting estimates of $H_0$, where the shaded envelope bracketing the
curves is the 95\% confidence region for the combined lens sample.  The
horizontal band shows the Key Project estimate.  For larger $f_{b,cold}$, the
density $\kbar$ decreases and the local slope $\eta$ steepens, leading to
larger values of $H_0$.  The vertical bands in the two panels show the lower
bound on $f_b$ from local inventories and the upper bound from the CMB.
\label{fig:cdm}}
\end{figure*}

There is an enormous range of parametric models that can illustrate
how the extent of the halo affects $\kbar$ and hence $H_0$ ---  the 
pseudo-Jaffe model we used above is only one example. It is
useful, however, to use a physically motivated model where the lens
galaxy is embedded in a standard NFW (Navarro, Frenk, \& White 1996) profile 
halo.  The lens galaxy consists of the baryons that have cooled to form stars, 
so the mass of the visible galaxy can be parameterized using the cold baryon
fraction $f_{b,cold}$ of the halo, and for these CDM halo models the value
of $\kbar$ is controlled by the cold baryon fraction (Kochanek~2003a).
A constant $M/L$ model is the limit 
$f_{b,cold} \rightarrow 1$ (with $\kbar \simeq 0.2$, $\eta \simeq 3$).
Since the baryonic mass fraction of a CDM halo should not exceed
the global fraction of $f_b \simeq 0.15 \pm0.05$ (e.g., Wang, 
Tegmark, \& Zaldarriaga 2002),
we cannot use constant $M/L$ models without also abandoning CDM.
As we reduce $f_{b,cold}$, we are adding mass to an extended 
halo around the lens, leading to an increase in $\kbar$ and a decrease
in $\eta$.  For $f_{b,cold} \simeq 0.02$ the model closely resembles
the SIS model ($\kbar\simeq 1/2$, $\eta \simeq 2$).  If we reduce $f_{b,cold}$
further, the mass distribution begins to approach that of the NFW halo
without any cold baryons. Figure~\ref{fig:cdm} shows how $\kbar$ and
$H_0$ depend on $f_{b,cold}$ for PG1115+080, SBS1520+530, B1600+434 and
HE2149--2745.  When $f_{b,cold} \simeq 0.02$, the CDM models have parameters
very similar to the SIS model, and we obtain a very similar estimate
of $H_0=52\pm6$~\kmsmpc\  (95\% confidence).  If all baryons cool, and 
$f_{b,cold}=f_b$, then we obtain $H_0=65\pm6$~\kmsmpc\  (95\% confidence), 
which is still lower than the Key Project estimates.    

We excluded the lenses requiring significantly
more complicated models with multiple lens galaxies or very strong
perturbations from clusters.  If we have yet
to reach a consensus on the mass distribution of relatively isolated
lenses, it seems premature to extend the discussion to still more
complicated systems.  We can, however, show that the clusters lenses
require significant contributions to $\kbar$ from the cluster in 
order to produce the same $H_0$ as the more isolated systems.
As we discussed in \S\ref{sec:data} the three more complex systems
are RXJ0911+0551, Q0957+561 and B1608+656.

RXJ0911+0551 is very strongly perturbed by the nearby X-ray cluster
(Morgan et al.~2001; Hjorth et al.~2002).  Kochanek (2003b) found
$H_0=49 \pm 5$~\kmsmpc\  if the primary lens and its satellite were 
isothermal and $H_0=67 \pm 5$~\kmsmpc\  if they had constant mass-to-light
ratios.  The higher value of   
$H_0=71\pm4$~\kmsmpc\  obtained by Hjorth et al.~(2002) can be
understood by combining \S\ref{sec:theory} and \S\ref{sec:clusters}
with the differences in the models.  In particular, Hjorth et al.~(2002)
truncated the halo of the primary lens near the Einstein radius
and used a lower mass cluster, both of which lower $\kbar$ and 
raise $H_0$.  The Hjorth et al.~(2002) models also included many
more cluster galaxies assuming fixed masses and halo sizes.

Q0957+561 is a special case because the primary lens galaxy is the
brightest cluster galaxy and it lies nearly at the cluster center
(Keeton et al.~2000; Chartas et al.~2002).  As a result, the 
lens modeling problems are particularly severe, and Keeton 
et al.~(2000) found that all previous models (most recently,
Barkana et al.~1999; Bernstein \& Fischer~1999; and Chae~1999) 
were incompatible with the observed geometry of the lensed host 
galaxy.  While Keeton et al.~(2000) found models consistent with
the structure of the lensed host, they covered a range of almost 
$\pm25\%$ in their estimates of $H_0$.  A satisfactory
treatment of this lens remains elusive.    

HE1104--1805 had its delay measured (Ofek \& Maoz 2003) just as we
completed this review.  Assuming the $\Delta t=161\pm7$~day delay is 
correct, a standard SIE model of this system predicts a very high 
$H_0 \simeq 90$~\kmsmpc.  The geometry of this system and the
fact that the inner image is brighter than the outer image both
suggest that HE1104--1805 lies in an anomalously high tidal shear
field, while the standard model includes a prior to keep the 
external shear small.  A prior is needed because a two-image lens 
supplies too few constraints to determine both the
ellipticity of the main lens and the external shear simultaneously.
Since the images and the lens in HE1104--1805 are nearly colinear,
the anomalous $H_0$ estimate for the standard model may be an
example of the shear degeneracy we briefly mentioned in \S\ref{sec:theory}.
At present the model surveys needed to understand the new delay
have not been made. Observations of the geometry of the host 
galaxy Einstein ring will resolve any ambiguities due to the shear
in the near future (see \S\ref{sec:profile}). 

The lens B1608+656 consists of two interacting galaxies, and, as
we discussed in \S\ref{sec:data}, this leads to a 
greatly increased parameter space.  Fassnacht et al.~(2002)
used SIE models for the two galaxies to find $H_0=61-65$~\kmsmpc,
depending on whether the lens galaxy positions are taken from the $H$-band or 
$I$-band lens {\it HST}\ images (the statistical errors are 
negligible).  The position differences are probably created by extinction
effects from the dust in the lens galaxies.
Like isothermal models of the ``simple'' lenses,
the $H_0$ estimate is below local values, but the disagreement is
smaller.  These models correctly match the observed time delay
ratios.  

\section{Solving the Central Concentration Problem \label{sec:profile} }

We can take four approaches to solving the central concentration problem.
First, the density profiles of galaxies are not a complete mystery, and
  we could apply the constraints derived from observations of other 
  (early-type) galaxies to the time delay systems.
Second, we could make new observations of the existing time delay lenses
  in order to obtain additional data that would constrain the density
  profiles.
Third, we could measure the time delays in the systems where the 
  lens galaxies already have well-constrained densities.
Fourth, we can use the statistical properties of time delay lenses to
  break the degeneracies seen in individual lenses.

If we assume that the time delay lenses have the same density structure as
other early-type galaxies, then models close to isothermal are favored.
For lenses with extended or multi-component sources, the lens models 
constrain the density distributions and the best fit models are usually 
very close to isothermal (e.g., Cohn et al. 2001; Winn, Rusin, \& Kochanek 2003).  Stellar 
dynamical observations of lenses also favor isothermal models (e.g., Treu \& 
Koopmans~2002a).  Stellar dynamical (e.g., Romanowsky \& Kochanek~1999; Gerhard et al.~2001)
and X-ray (e.g., Loewenstein \& Mushotzky~2003) observations of nearby early-type 
galaxies generally find flat rotation curves on the relevant scales.
Finally, weak lensing analyses
require significant dark matter on large scales in early-type galaxies
(McKay et al.~2002).  In general, the data on early-type galaxies seem to 
prefer isothermal models on the scales relevant to interpreting time delays,
while constant $M/L$ models are firmly ruled out.  If we must ultimately
rely on the assumption that the density profiles of time delay lenses
are similar to those of other early-type galaxies, the additional uncertainty
added by this assumption will be small and calculable. Moreover, the 
assumption is no different from the assumptions of homogeneity used
in other studies of the distance scale.
   
We can avoid any such assumptions by determining
the density profiles
of the time delay lenses directly.  One approach is to measure the
kinematic properties of the lens galaxy.  Since the mass 
inside the Einstein ring is fixed by the image geometry, the velocity
dispersion is controlled by the central concentration of the density.
Treu \& Koopmans (2002b) apply this method to  PG1115+080 and
argue that the observed velocity dispersion 
requires a mass distribution between the isothermal and constant
$M/L$ limits with $H_0=59^{+12}_{-7}$~\kmsmpc. 
 Note, however, that with this velocity dispersion the
lens galaxy does not lie on the fundamental plane, which is very 
peculiar.    A second approach is to use deep infrared imaging to
determine the structure of the lensed host galaxy of the quasar 
(Kochanek, Keeton, \& McLeod 2001).  The location and width of the Einstein ring 
depends on both the radial and angular structure of the potential,
although the sensitivity to the radial structure of the lens
is weak when the annulus bracketing the lensed images is thin
($\dr/\rbar$ small; Saha \& Williams~2001).  This method will
work best for asymmetric two-image lenses ($\dr/\rbar \approx 1$).
The necessary data can be obtained with {\it HST}\ for most time delay 
lenses.

We can also focus our monitoring campaigns on lenses already known to have 
well-constrained density profiles.  For the reasons we have already 
discussed, systems with multi-component sources, well-studied images
of the host galaxy or stellar dynamical measurements will have better
constrained density profiles than those without any additional constraints.
We can also avoid most of the uncertainties in the density profile by
measuring the time delays of very low-redshift lenses.  When the lens
is very close to the observer, the images lie very close to the center
of the lens where the stellar mass dominates. A constant $M/L$ model
then becomes a very good approximation and we need worry little about
the amount or the distribution of the dark matter.  The one such 
candidate at present, Q2237+0305 at $z_l=0.04$, will have very short
delays, but these could be measured by an X-ray monitoring program 
using the {\it Chandra} observatory.

Finally, the statistical properties of larger samples of time delay lenses
will also help to solve the problem.  We already saw in \S\ref{sec:results}
that the ``simple'' time delay lenses must have very similar densities,
independent of $H_0$.  This already means that the implications for $H_0$
no longer depend on individual lenses.  In some ways the similarity of
the densities is not an advantage --- it is actually easier to determine
$H_0$ if the density distributions are inhomogeneous (Kochanek~2003b). 
On the other hand, there are well-defined approaches to using the statistical
properties of lens models to estimate parameters that cannot be determined
from the models of the individual systems (see Kochanek~2001).  
The statistics of the problematic flux ratios 
observed in the lenses (see \S\ref{sec:data}) may also 
provide a means of estimating $\kbar$.  Schechter \& Wambsganss~(2002)
point out that in four-image quasar lenses there is a tendency for the
brightest saddle point image to be demagnified compared to reasonable
lens models.  Microlensing by the stars can naturally explain the 
observations if the surface density of stars is a small fraction of
the total surface density near the images ($\kappa_* \ll \kbar$),
which would rule out constant $M/L$ models where $\kappa_* \simeq \kbar$.

\section{Conclusions \label{sec:future} }

The determination of $H_0$ using gravitational lens time delays has come
of age. The last few years have seen a dramatic increase in the number
of delay measurements, and there is no barrier other than sociology to
rapidly increasing the sample.  The interpretation of time delays requires a model for
the gravitational potential of the lens.  Fortunately, the physics
determining time delays is well understood, and the only important
variable is the average surface density $\kbar$ of the lens near the 
images for which the delay is measured.  Unfortunately, there is a tendency 
in the literature to conceal rather than to illuminate this understanding.
Provided a lens does not lie in a cluster where the cluster potential cannot 
be described by a simple expansion, any lens model that includes the parameters 
needed to vary the average surface density of the lens near the images and
to change the ratio between the quadrupole moment of the lens and the 
environment includes all the parameters needed to model time delays,
to estimate the Hubble constant, and to understand the systematic 
uncertainties in the results. {\it All differences between estimates of the
Hubble constant for the simple time delay lenses can be understood on
this basis.}

Models for the four time delay lenses that can be modeled using a single
lens galaxy predict that $H_0=\vsis$~\kmsmpc\  if the lens galaxies have 
isothermal density profiles with flat rotation curves, and $H_0=\vml$~\kmsmpc\ 
if they have constant mass-to-light ratios.  The Key Project estimate of 
$H_0=72\pm8$~\kmsmpc\  agrees with the lensing results only if the lenses 
have little dark matter.  We have strong theoretical prejudices and estimates 
from other observations of early-type galaxies that we should favor the 
isothermal models over the constant $M/L$ models.  We feel that we
have reached the point where the results from gravitational lens time
delays deserve serious attention and that there is a reasonable likelihood
that the local estimates of $H_0$ are too high.  A modest investment 
of telescope time would allow the measurement of roughly 5--10 time delays
per year, and these new delays would rapidly test the current results. 
Other observations of time
delay lenses to measure the velocity dispersions of the lens galaxies
or to determine the geometry of the lensed images of the quasar host
galaxy can be used to constrain the mass distributions directly.  
The systematic problems associated with the density profile are soluble
not only in theory but also in practice, and the investment of the
community's resources would be significantly less that than already
invested in the distance scale.
 
The time delay measurements also provide a new probe of the density 
structure of galaxies at the boundary between the baryonic and dark matter
dominated parts of galaxies (projected distances of 1--2 effective
radii). Even if we ignore the actual value of $H_0$, we can still
study the differences in the surface densities.  For example, we
can show that the present sample of simple lenses must have very
similar surface densities.  This region is very difficult to study
with other probes.    

Finally, the time delay measurements can be used to determine cosmological
parameters.  Time delays basically measure the distance to the lens galaxy,
so we can make the same sorts of cosmological measurements
as Type Ia supernovae.  If the variations in $\kbar$ between lens galaxies
are small, as seem to be indicated by the present data, then the accuracy
of the differential measurements will be very good.  The present sample
has little sensitivity to the cosmological model even with the mass 
distribution fixed because the time delay uncertainties are still too 
large and the redshift range is too restricted ($z_l=0.31$ to $0.72$).  
If we assume that other methods will determine the distance factors 
more accurately and rapidly, then we can use the time delays to study
the evolution of galaxy mass distributions with redshift.  

\vspace{0.3cm}
{\bf Acknowledgements}.
CSK thanks D. Rusin and J. Winn for their comments.  CSK is supported by the 
Smithsonian Institution and NASA ATP grant NAG5-9265.


\begin{thereferences}{}

\bibitem{}
Angonin-Willaime, M.-C., Soucail, G., \& Vanderriest, C. 1994, A\&A, 291, 411

\bibitem{}
Barkana, R. 1996, ApJ, 468, 17

\bibitem{}
------. 1997, ApJ, 489, 21

\bibitem{}
Barkana, R., Leh\'ar, J., Falco, E.~E., Grogin, N. A., Keeton, C. R., \& 
Shapiro, I.~I. 1999, ApJ, 520, 479

\bibitem{}
Bernstein, G., \& Fischer, P. 1999, AJ, 118, 14 

\bibitem{}
Beskin, G. M., \& Oknyanskij, V. L. 1995, A\&A, 304, 341

\bibitem{}
Biggs, A. D., Browne, I. W. A., Helbig, P., Koopmans, L. V. E., Wilkinson, P. 
N., \& Perley, R. A. 1999, MNRAS, 304, 349

\bibitem{}
Blandford, R. D., \& Narayan, R. 1987, ApJ, 310, 568

\bibitem{}
Bullock, J.~S., Kolatt, T.~S., Sigad, Y., Somerville, R.~S., Kravtsov,
A.~V., Klypin, A.~A., Primack, J.~R., \& Dekel, A. 2001, MNRAS, 321, 559

\bibitem{}
Burud, I., et al. 2000, ApJ, 544, 117

\bibitem{}
------. 2002a, A\&A, 383, 71

\bibitem{}
------. 2002b, A\&A, 391, 481

\bibitem{}
Chae, K.-H. 1999, ApJ, 524, 582

\bibitem{}
Chartas, G. 2003, Carnegie Observatories Astrophysics Series, Vol. 2:
Measuring and Modeling the Universe, ed. W. L. Freedman
(Pasadena: Carnegie Observatories,
http://www.ociw.edu/ociw/symposia/series/symposium2/proceedings.html)

\bibitem{}
Chartas, G., Gupta, V., Garmire, G., Jones, C., Falco, E. E., Shapiro, I. I., 
\& Tavecchio, F. 2002, ApJ, 565, 96

\bibitem{}
Cohen, A. S., Hewitt, J. N., Moore, C. B., \& Haarsma, D. B. 2000, ApJ, 545, 578

\bibitem{}
Cohn, J. D., Kochanek, C. S., McLeod, B. A., \& Keeton, C. R. 2001, 
ApJ, 554, 1216

\bibitem{}
Courbin, F., Meylan, G., Kneib, J.-P., \& Lidman, C. 2002, ApJ, 575, 95

\bibitem{}
Dalal, N., \& Kochanek, C. S. 2002, ApJ, 572, 25

\bibitem{}
Fassnacht, C. D, Xanthopoulos, E., Koopmans, L. V. E., \& Rusin, D. 2002, ApJ,
581, 823

\bibitem{}
Falco, E. E., Gorenstein, M. V., \& Shapiro, I. I. 1985, ApJ, 289, L1

\bibitem{}
Fischer, P., Bernstein, G., Rhee, G., \& Tyson, J. A. 1997, AJ, 113, 521 

\bibitem{}
Florentin-Nielsen, R. 1984, A\&A, 138, L19

\bibitem{}
Freedman, W. L., et al. 2001, ApJ, 553, 47


\bibitem{}
Gerhard, O., Kronawitter, A., Saglia, R. P., \& Bender, R. 2001, AJ, 121, 1936

\bibitem{}
Gil-Merino, R., Wisotzki, L., \& Wambsganss, J. 2002, A\&A, 381, 428

\bibitem{}
Gorenstein, M. V., Falco, E. E., \& Shapiro, I. I. 1988, ApJ, 327, 693

\bibitem{}
Haarsma, D. B., Hewitt, J. N., Leh\'ar, J.,  \& Burke, B. F. 1999, AJ, 510, 64

\bibitem{}
Hjorth, J., et al., 2002, ApJ, 572, L11

\bibitem{}
Impey, C. D., Falco, E. E., Kochanek, C. S., Leh\'ar, J., McLeod, B. A., 
Rix, H.-W., Peng, C. Y., \& Keeton, C. R.  1998, ApJ, 509, 551

\bibitem{}
Keeton, C. R., et al. 2000, ApJ, 542, 74

\bibitem{}
Keeton, C. R. 2003, ApJ, submitted (astro-ph/0102340)

\bibitem{}
Keeton, C. R., Kochanek, C. S., \& Seljak, U. 1997, ApJ, 482, 604

\bibitem{}
Kochanek, C. S. 2001, in The Shapes of Galaxies and Their Dark Halos, 
ed. P. Natarajan (Singapore: World Scientific), 62

\bibitem{}
------. 2002, ApJ, 578, 25

\bibitem{}
------. 2003a, ApJ, 583, 49

\bibitem{}
------. 2003b, astro-ph/0204043

\bibitem{}
Kochanek, C. S., Keeton, C. R., \& McLeod, B. A. 2001, ApJ, 547, 50

\bibitem{}
Koopmans, L. V. E., de Bruyn, A. G., Xanthopoulos, E., \& Fassnacht, C. D. 
2000, A\&A, 356, 391

\bibitem{}
Kundi\'c, T., et al. 1997, ApJ, 482, 75

\bibitem{}
Leh\'ar, J., et al. 2000, ApJ, 536, 584

\bibitem{}
Lovell, J. E. J., Jauncey, D. L., Reynolds, J. E., Wieringa, M. H., King, 
E. A., Tzioumis, A. K., McCulloch, P. M., \& Edwards, P. G. 1998, ApJ, 508, L51

\bibitem{}
Loewenstein, M., \& Mushotzky, R. F. 2003, ApJ, submitted 

\bibitem{}
McKay, T. A., et al. 2002, ApJ, 571, L85

\bibitem{}
Metcalf, R. B., \& Madau, P. 2001, ApJ, 563, 9

\bibitem{}
Morgan, N. D., Chartas, G., Malm, M., Bautz, M. W., Burud, I., Hjorth, J., 
Jones, S. E., \& Schechter, P. L. 2001, ApJ, 555, 1 

\bibitem{}
Narayan, R., \& Bartelmann, M. 1999, in Formation and Structure in the 
Universe, ed. A. Dekel \& J. P. Ostriker (Cambridge: Cambridge Univ. Press), 360

\bibitem{}
Navarro, J. F., Frenk, C. S., \& White, S. D. M. 1996, ApJ, 462, 563

\bibitem{}
Ofek, E. O., \& Maoz, D. 2003, ApJ, in press 

\bibitem{}
Patnaik, A. R., \& Narasimha, D. 2001, MNRAS, 326, 1403

\bibitem{}
Pelt, J., Kayser, R., Refsdal, S., \& Schramm, T. 1996, A\&A, 305, 97

\bibitem{}
Pelt, J., Refsdal, S., \& Stabell, R. 2002, A\&A, 289, L57

\bibitem{}
Press, W. H., Rybicki, G. B., \& Hewitt, J. N. 1992a, ApJ, 385, 404

\bibitem{}
------. 1992b, ApJ, 385, 416

\bibitem{}
Refsdal, S. 1964, 128, 307

\bibitem{}
Romanowsky, A. J., \& Kochanek, C. S. 1999, ApJ, 516, 18 

\bibitem{}
Saha, P. 2000, AJ, 120, 1654

\bibitem{}
Saha, P., \& Williams, L. L. R. 2001, AJ, 122, 585

\bibitem{}
Schechter, P. L., et al. 1997, ApJ, 475, L85 

\bibitem{}
------. 2003, ApJ, 584, 657 

\bibitem{}
Schechter, P. L., \& Wambsganss, J. 2002, ApJ, 580, 685

\bibitem{}
Schild, R., \& Cholfin, B. 1986, ApJ, 300, 209

\bibitem{}
Schild, R., \& Thomson, D. J. 1997, AJ, 113, 130

\bibitem{}
Schneider, P., Ehlers, J., \& Falco, E. E. 1992, Gravitational Lenses,
(Berlin: Springer-Verlag) 

\bibitem{}
Treu, T., \& Koopmans, L. V. E. 2002a, ApJ, 568, L5

\bibitem{}
------. 2002b, MNRAS, 337, 6

\bibitem{}
Vanderriest, C., Schneider, J., Herpe, G., Chevreton, M., Moles, M., \& 
Wlerick, G. 1989, A\&A, 215, 1 

\bibitem{}
Wang, X., Tegmark, M., \& Zaldarriaga, M. 2002, Phys. Rev. D, 65, 123001

\bibitem{}
Williams, L. L. R., \& Saha, P. 2000, AJ, 199, 439

\bibitem{}
Winn, J. N., Kochanek, C. S., McLeod, B. A., Falco, E. E., Impey, C. D., \& 
Rix, H.-W. 2002, ApJ, 575, 103  

\bibitem{}
Winn, J. N., Rusin, D., \& Kochanek, C. S. 2003, ApJ, 587, 80

\bibitem{}
Witt, H. J., Mao, S., \& Keeton, C. R. 2000, ApJ, 544, 98

\bibitem{}
Wozniak, P. R., Udalski, A., Szymanski, M., Kubiak, M., Pietrzynski, G., 
Soszynski, I., \& Zebrun, K. 2000, ApJ, 540, L65 

\bibitem{}
Zhao, H., \& Qin, B. 2003a, ApJ, 582, 2

\bibitem{}
------. 2003b, ApJ, submitted (astro-ph/0209304)

\end{thereferences}
\end{document}